\documentclass{aa}  

\usepackage{graphicx}
\usepackage{txfonts}
\usepackage{natbib}
\usepackage{xspace}
\usepackage[pdftitle={M.Cardillo-Orion-V1}, colorlinks=true, citecolor=blue, linkcolor=blue, urlcolor=blue, pdfauthor={Martina Cardillo}]{hyperref}

\def \gray {$\gamma$-ray\xspace}

\def \kori {$\kappa$-Ori\xspace}
\def \dg {\textquotedbl dark gas\textquotedbl \xspace}

\begin{document}

%\bibliographystyle{apj}
%\submitted{To be sumbmitted to \textit{A$\&$A}. Draft n.01 \today}

\title{The Orion Region: Evidence of enhanced cosmic-ray density in a stellar wind forward shock interaction with a high density shell}

\author{M.~Cardillo\inst{1}
\and N.~Marchili\inst{1} \and G.~Piano\inst{1} \and A.~Giuliani\inst{5} \and M.~Tavani\inst{1,2,3,4} \and S.~Molinari\inst{1}}

\institute{INAF-IAPS, Via del Fosso del Cavaliere 100 - 00133 Roma, Italy\\
e-mail:martina.cardillo@inaf.it
\and Dip. di Fisica, Univ. Tor Vergata, Via della Ricerca Scientifica 1, I-00133 Roma, Italy
\and Gran Sasso Science Institute, viale Francesco Crispi 7, I-67100 L'Aquila, Italy
\and Astronomia, Accademia Nazionale dei Lincei, Via della Lungara 10, I-00165 Roma, Italy
\and INAF - IASF Milano, Via E. Bassini 15, I-20133, Milano, Italy.}

\date{Received / Accepted }

%****************************************abstract
\abstract{
\textit{Context}. In recent years, an in-depth \gray analysis of the Orion region has been carried out by the AGILE and Fermi-LAT (Large Area Telescope) teams with the aim of estimating the H$_2$-CO conversion factor, $X_{CO}$. The comparison of the data from both satellites with models of diffuse \gray Galactic emission unveiled an excess at (l,b)=[213.9, -19.5], in a region at a short angular distance from the OB star \kori. Possible explanations of this excess are scattering of the so-called \dg, non-linearity in the  H$_2$-CO relation, or Cosmic-Ray (CR) energization at the \kori wind shock.\\
\textit{Aims}. Concerning this last hypothesis, we want to verify whether cosmic-ray acceleration or re-acceleration could be triggered at the k-Ori forward shock, which we suppose to be interacting with a star-forming shell detected in several wavebands and probably triggered by high energy particles.\\
\textit{Methods}. Starting from the AGILE spectrum of the detected \gray excess, showed here for the first time, we developed a valid physical model for cosmic-ray energization, taking into account re-acceleration, acceleration, energy losses, and secondary electron contribution.\\
\textit{Results}. Despite the characteristic low velocity of an OB star forward shock during its 'snowplow' expansion phase, we find that the Orion \gray excess could be explained by re-acceleration of pre-existing cosmic rays in the interaction between the forward shock of \kori and the CO-detected, star-forming shell swept-up by the star expansion. According to our calculations, a possible contribution from freshly accelerated particles is sub-dominant with respect the re-acceleration contribution. However, a simple adiabatic compression of the shell could also explain the detected \gray emission. Futher GeV and TeV observations of this region are highly recommended in order to correctly identify the real physical scenario.

}

\keywords{Acceleration of particles, Radiation mechanisms: non-thermal, Methods: data analysis, Stars: winds, Gamma rays: ISM, ISM:cosmic rays}

\titlerunning{Orion}
\authorrunning{M.Cardillo}

\maketitle

%%%%%%%%%%%%%%%%%%%%%%%%%%%%%%%%%%%%%%%
\section{Introduction}
\label{sec:intro}
%%%%%%%%%%%%%%%%%%%%%%%%%%%%%%%%%%%%%%% 
The Orion region and its surroundings (\citealt{Bally08}, and references therein) is amongst the most studied Galactic region because it is the nearest star formation site known so far, and it is less affected by Galactic diffuse emission. It includes different structures: the stellar association Orion OB1, with about one hundred OB stars, and two giant molecular clouds (MCs), Orion A and Orion B \citep{Pillitteri16}.

In recent years, an important survey by the X-ray satellite XMM-Newton in the Orion region has revealed the presence of several Young Stellar Objects (YSO) in a high density shell overlapping the MC Orion A but probably unrelated to it \citep{Pillitteri16}. This shell is shaped as a ring of 5-8 pc radius around \kori, a blue super giant B0.5 Ia star at a distance of 240-280 pc from us, with an estimated age of 7x10$^6$ yrs. Detections in the near- and far-IR, especially in the CO band, together with the analysis of new XMM-Newton data, show evidence that this shell is closer than the Orion A MC, at a distance compatible with that of \kori. Consequently, the star formation process within it could have been triggered by the \kori wind expansion that has swept up the shell, a hypothesis compatible with the estimated dynamical age of about $1.5$ Myr \citep{Wilson05}, which is lower by about one order of magnitude than the \kori age.

At the highest energies, \gray emission from the Orion region was firstly detected by the COS-B (Cosmic ray Satellite ('option B')) \citep{Caraveo80} and EGRET \citep{Digel95} satellites. Recent observations by Fermi-LAT \citep{Ackermann12} and AGILE \citep[][hereafter M18]{Marchili18} led to a more precise characterization of the diffuse emission from this region, allowing for a comparison with the expected flux from standard models. These take generally into account Bremsstrahlung and proton-proton (pp) emission related to the HI and H$_2$ distributions, inverse-Compton scattering on the InterStellar Radiation Field (ISRF), or CMB photons, an isotropic component due to the extragalactic emission and the contribution of point-like or extended known \gray sources. The analysis of Fermi-LAT and AGILE data shows significant emission exceeding the estimations of standard models, in a location overlapping the high-longitude part of the Orion A MC. A first attempt to explain this excess was done by adding a \dg (gas not traced by HI and CO) contribution to diffuse \gray emission, following the approach introduced in \cite{Grenier05}. However, both Fermi-LAT and AGILE data analysis shows that this has only a marginal effect on the detected excess; therefore, another kind of physical mechanism is required to explain it. In \cite{Ackermann12}, in the light of some recent studies (see references within their paper), the authors hypothesized a non-linearity in the CO-H$_2$ relation that implies a conversion factor, $X_{CO  }$, which is not constant. In M18, we introduced an alternative hypothesis. The AGILE-detected \gray excess seems to be in a good correlation with a star formation shell observed in the velocity maps from CO surveys and confirmed by X-ray observations of the XMM-Newton satellite, analysed in \cite{Pillitteri16}.

In this context, we considered the possibility of CR acceleration at the shock where the stellar wind collides with the InterStellar Medium (ISM) (M18). Previous studies (e.g. \citealt{Casse80}, \citealt{Voelk82}, \citealt{Cesarsky83}, \citealt{Ip95}) suggested that the ideal location for CR acceleration is the strong Termination Shock (TS) of a stellar wind, because of its high velocity (order of $10^2-10^3$ km/s). We consider instead the possibility that CR energization occurs at the Forward Shock (FS) of \kori, despite its relatively slow velocity \citep[especially in the `snowplow' expansion phase, which describes well the present state of \kori; see][]{Lamers99}. The detection in the CO of the high density star-forming shell around \kori, partially overlapping the excess emission detected by AGILE, strongly supports our hypothesis.\\

In this paper, we discuss the AGILE \gray spectrum of the detected excess, assuming that it originates in the inner part of the star-forming shell described in \cite{Pillitteri16}. Following both acceleration and re-acceleration model described in \cite{Cardillo16} (hereafter Ca16), we try to explain the AGILE-detected emission in the context of CR re-acceleration and/or acceleration, taking into account all the known parameters and giving an estimation of the average density in that region. In Sect. \ref{Sec:OrionGamma}, we summarize the results presented in M18, showing for the first time the AGILE spectral points. In Sect. \ref{Sec:Model}, we give an overview of the model used in order to fit AGILE data in the context of CR re-acceleration and acceleration based on Ca16. In Sect. \ref{Sec:Results}, we show our best results, and in Sect. \ref{Sec:Discussions} we analyse the physical consequences of all models and discuss different assumptions and parameters. The main points of the whole work are summarized in Sect. \ref{Sec:Conclusions}.

%%%%%%%%%%%%%%%%%%%%%%%%%%%%%%%%%%%%%%%
\section{Orion \gray excess}
\label{Sec:OrionGamma}
%%%%%%%%%%%%%%%%%%%%%%%%%%%%%%%%%%%%%%% 

In M18, we analysed AGILE data in the well-known Orion region with the aim of establishing a deeper understanding of the nature of its \gray emission. The large abundance of ISM suggests that diffuse \gray emission should be very abundant in that site. We focused on a region of interest (ROI) of $15^{\circ}\times 11^{\circ}$ centred at (l,b)=[210.5,-15.5] (M18) and modelled \gray diffuse emission in different steps. In the first, we considered the Galactic contribution due to pp-interactions and Bremsstrahlung emissivity per Hydrogen atom (HI and $H_{2}$), and to Inverse Compton (IC) emissivity from the ISRF, removing distance and energy dependences through spatial and energetic integration. Then, we included the extra-galactic contribution as an isotropic component, neglecting the contribution from point sources because the AGILE-GRID source catalogue does not contain any object inside the ROI. Finally, we fitted the AGILE total \gray flux $S_{tot}$ as a linear combination of the contributions due to HI ($S_{HI}$), H$_{2}$ ($S_{H_{2}}$), and the whole isotropic component (IC and extragalactic contributions, added up in the variable $\epsilon$):
%*********************
\begin{equation}
 S_{tot}(l,b)=\alpha S_{HI}(l,b)+\beta S_{H_{2}}(l,b)+\epsilon
 \label{eq:AGILE_totFlux}
.\end{equation}
%*********************
In this equation, $\alpha$ and $\beta$ represent the \gray emissivity due to pp and Bremsstrahlung emission, therefore, if we have computed in the correct way the spatial distribution of HI and H$_2$, these coefficients should be equal (M18). A difference in their values would be due to the uncertainty in the column density estimators. We can provide an estimation of the effective CO-H$_{2}$ conversion factor, $\left(\mathrm{X}_\mathrm{CO}\right)_{eff}$, which is correlated with the real conversion factor through $\alpha$ and $\beta$ ratio
%********************
\begin{equation}
 \left(\mathrm{X}_\mathrm{CO}\right)_{eff}=\frac{\beta}{\alpha}\mathrm{X}_\mathrm{CO}=(1.32\pm0.05)\times10^{20} \textit{$cm^{-2}K^{-1}km^{-1}$}
 \label{eq:conv_coeff}
,\end{equation}
%********************
taking into account possible uncertainties in the column density estimation. The value obtained is in very good agreement with previous measurements by EGRET \citep{Digel99} and Fermi-LAT \citep{Ackermann12}. Moreover, the AGILE residual map shows a further confirmation of the Fermi result: we found a \gray excess accounted for neither by atomic and/or molecular hydrogen nor % Fuori dal piano galattico la componente molecolare e' decisamente dominante
by the extragalactic component;  this excess is centred at (l,b)=[213.9, -19.5] and perfectly overlaps an excess found in \cite{Ackermann12}. It is arc-shaped and seems to belong to a ring centred on the position of the B0.5 Ia star \kori, about $4$ pc away from it. \cite{Ackermann12} try to explain the detected \gray excess with the presence of \dg, a gas component not traced by HI or CO % per coerenza, se ti riferisci ad un atomo/molecola (nella fattispecie HI) come tracciatore, e' opportuno affiancargli CO piuttosto che W_CO
but possibly by thermal IR emission \citep{Grenier05}. After introducing this \dg contribution to the emission, which they computed following a template based on reddening map of IRAS (InfraRed Astronomical Satellite) and COBE (COsmic Background Explorer) \citep{Schlegel98}, the diffuse emission model of the whole Orion region provides a better description of the observed flux; the improvement, however, is much more significant for outer regions than for Orion A. A better modelling of the Orion A \gray excess is achieved by assuming a non-linear behaviour of the CO$-$H$_2$ relation, which implies a variable conversion factor. This hypothesis is in line with several works developed in the last years \citep[see][and references therein]{Ackermann12}, according to which, in sites with high star formation rates, the ratio between the two molecules is variable and depends on environmental parameters.

AGILE extended likelihood analysis estimates a flux of $(11 \pm 2)\times 10^{-8}$ $ph/cm^{2}/s$ with a significance of $5.2\sigma$. In our previous work, we considered three different templates for the computation of the \dg contribution (for details see M18), amongst which also the one used in \cite{Ackermann12}. The best-fit template for \dg appeared to be the reddening map derived from the Pan-STARRS1 (Panoramic Survey Telescope $\&$ Rapid Response System) stellar photometry in \cite{Schlafly14}; we used it as an estimate of the \dg contribution to the total flux computation in Eq. \ref{eq:AGILE_totFlux}. As in \cite{Ackermann12}, the new model improves the description of the observed diffuse emission; the \dg template, however, seems to account for only  $\sim25\%$ of the \gray excess detected in the high-longitude part of Orion A.

\begin{figure}[!h]
\begin{center}
\includegraphics[scale=0.68]{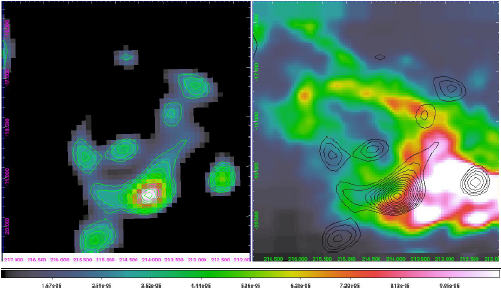}
 % AGILE_shell.png: 0x0 pixel, 0dpi, nanxnan cm, bb=
\end{center}
\caption{\textbf{Left}: Ring of \gray excess detected by AGILE. \textbf{Right}: CO map \citep{Dame01}, which reveals the star formation shell discussed by \cite{Pillitteri16}; the contour levels from \gray data are shown in black. The figure is from M18.}
\label{Fig:AGILEem}
\end{figure}

Looking for an alternative explanation, we considered the results of \cite{Pillitteri16}. Analysing data obtained from a XMM-Newton X-ray survey, they associated an arc-shaped emission detected in the IR, X-ray, and CO bands to a star formation region that they locate at a distance between 5 and 8 pc from \kori. Looking at Fig.\ref{Fig:AGILEem}, it is evident that the AGILE \gray emission arises in the inner part of this shell, partially overlapping the CO emission, likely in correspondence with the interaction of the FS of \kori 's strong stellar wind with the high density shell.

%**************************
%**************************

Given the striking similarities in shape and position between the \gray emission and the star formation shell in \cite{Pillitteri16}, we formulated a possible explanation of the \gray excess in the context of {\it in situ} CR acceleration and re-acceleration. Moreover, the AGILE extended source analysis of the excess gives a very hard spectral index, $1.7\pm0.2$, which could support a Diffusive Shock Acceleration (DSA) process \citep{Blasi04}, even if only a detailed theoretical model can give the real range of the parent proton population spectral index.

\begin{figure}[!h]
\begin{center}
 \includegraphics[scale=0.45]{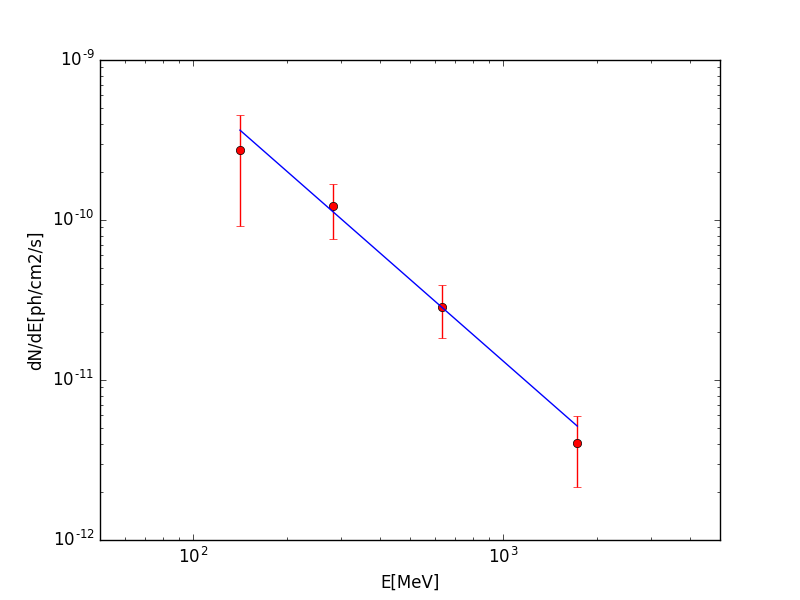}
 % orion_spectrum.eps: 0x0 pixel, 300dpi, 0.00x0.00 cm, bb=
\end{center}
 \caption{AGILE \gray spectrum obtained with an extended source likelihood analysis centred on (l, b) = (214.4, -18.5) with a $1^{\circ}$ radius convoluted with AGILE PSF (M18). The blue line shows the best-fit power-law function, with index 1.7.}
 \label{fig:spectrum}
\end{figure}

In Fig. \ref{fig:spectrum}, we show the \gray spectrum of the AGILE detected excess. The spectral points are obtained by analysing the AGILE data between November 2009 and March 2017, corresponding to AGILE spinning mode data-taking. We carried out an extended-source analysis by assuming, as template for \gray emission, a uniform disc centred on (l,b) = [214.4, -18.5] with a $1^{\circ}$ radius (the same used in M18). In order to perform a spectral analysis, the circle has been convoluted with the AGILE Point Spread Function (PSF) in each energy band. For this study, we used the \verb+Build 23+ AGILE software on the consolidated data archive, and the \verb+I0025+ response matrices. 
According to the spectral energy distribution estimated with AGILE, this \gray excess could be the first evidence of accelerated/re-accelerated CR presence in correspondence with a strong stellar wind of a single OB star.

%**************************
%**************************

%%%%%%%%%%%%%%%%%%%%%%%%%%%%%%%%%%%%%%%
\section{Model overview}
\label{Sec:Model}
%%%%%%%%%%%%%%%%%%%%%%%%%%%%%%%%%%%%%%%

The energetics of an OB star with an average wind terminal velocity $V_{w}\sim2\times10^{3}$ km/s and a mass loss rate between $10^{-7}-10^{-5}M_{\odot}/yr$ is between $10^{50}$ and $10^{51}$ erg, of the same order of the total SuperNova(SN) explosion energy \citep{Lamers99, Casse80, Cesarsky83}. The possibility of particle acceleration in correspondence with the TS of strong stellar winds was proposed in order to explain \gray emission detected by the Cos-B satellite from the $\rho$-Oph cloud \citep{Casse80,Voelk82, Cesarsky83}, and confirmed in recent years by the observations of \gray sources associated to young stellar clusters \citep[see, for instance]{HESS11}. However, AGILE detected \gray excess only in the south-west side of the \kori expanding wind. Even if this asymmetric \gray emission could be explained in terms of TS energization because of the presence of the CO shell only in that region, the FS interacts with it before the TS. Consequently we considered particle energization in correspondence with the interaction of the FS with the dense molecular shell. The OB star \kori is now in its `snowplow' phase \citep{Lamers99} and, consequently, the FS velocity should be of the order of a few tens of km/s. However, considering that the upstream is a cold ($T<100$ K) dense shell, the Mach number of the shock is of the order of 10 and we can consider the FS still a strong shock. Moreover, the estimated average density $n_{0}=30$ cm$^{-3}$ is large enough to trigger the formation of a radiative thin shell in the interaction location, which provides a very high compressed density and, consequently, enhances the \gray emission \citep[see Eq. 1 in][]{Blandford82}.

The scenario we propose to explain the cosmic-ray energization is similar to the typical one found in a SuperNova Remnant(SNR) shock, for example in the W44 SNR \citep{GiuCaTa11,Cardillo14,Ackermann13,Cardillo16}. Consequently, we model this emission following the approach described in Ca16; we take into account both re-acceleration and acceleration contributions from protons, helium, and electrons (primaries and secondaries) with their radiation losses. In M18, we estimated for our ROI a \gray flux $F_{\gamma}=11 \pm 10^{-8}$ ph/cm$^{2}$/s$^{-1}$, corresponding to a \gray luminosity between [100-3000] MeV, $L_{\gamma}=5.2\times10^{31}$ erg/s. Assuming that all the \gray luminosity is produced by pp interaction $\pi^{0}$ decay, we can estimate a total CR energy of $W_{CR}\sim4\times10^{45}$ erg \citep[see for instance][]{Aharonian18}. In the following, we summarize the main points of this model and the related physical issues.

The effect of re-acceleration is to harden the spectrum of Galactic particles with a spectrum steeper than $f(p)\propto p^{-\alpha}$ , where $\alpha$ is the DSA spectral index related to the compression ratio, that is, the ratio between upstream velocity (ahead of the shock) and downstream velocity (behind the shock), $r_{sh}=\frac{v_{up}}{v_{down}}$. In the strong shock approximation \citep{Amato14}, $\alpha=\frac{3r_{sh}}{r_{sh}-1}$. The re-accelerated particle spectrum is obtained with the  equation
%***************
\begin{equation}
f_{0}(p)=\alpha\left(\frac{p}{p_{m}}\right)^{-\alpha}\int^{p}_{p_{m}}\frac{dp'}{p'}\left(\frac{p'}{p_{m}}\right)^{\alpha}f_{\infty}(p')
\label{eq:Re}
,\end{equation}
%***************
where $p$ is the particle momentum and $p_{m}$ represents a minimum momentum in the Galactic CR spectrum, in this case equal to $1$ MeV, consistent with the lowest energy of the Voyager probe's spectral data, which is a few MeV. The distribution function $f_{\infty}(p')$ describes the Galactic CR spectrum for different kinds of particles (protons, electrons, and helium nuclei) that we modelled according to the Local InterStellar (LIS) spectrum measured by Voyager 1 at low energies ($E\gtrsim 1$ MeV/n; \citealt{Webber13_Voyager}), and PAMELA (Payload for Antimatter Matter Exploration and Light-nuclei Astrophysics) \citep{Adriani11_p} and AMS-02 (Alpha Magnetic Spectrometer) data \citep{Aguilar15_e,Aguilar15_p,Aguilar15_He} at higher energies.

%\textbf{Following the approach of \citealt{Voelk82}, we assume that energizated particles come from the interface between the shocked stellar wind and the swept up ambient gas. Here, Galactic CR spectrum will be different from the LIS spectrum used in Ca16 at energies below $30$ GeV/n. There,  for protons, helium and electrons LIS spectra. Here, instead,as in the solar wind, it is affected by modulation; consequently, we used a Galactic CR spectra obtained from the parametric equations in Ca16 bur taking into accunt only the high energy data, down to energies of about $0.5$ GeV/n.}
The distribution of possible freshly accelerated CRs (both hadrons and leptons) is described by the conservative power-law spectrum directly provided by DSA theory,
%******************
\begin{equation}
f_{i}(p)=k_{i}\left(\frac{p}{p_{inj}}\right)^{-\alpha}
\label{eq:acc}
,\end{equation}
%******************
where index $i$ reads $p$ for protons and $e$ for electrons. The normalization factor $k_{i}$ is obtained considering that CR pressure is a fraction $\xi_{CR}$ of ram pressure of the shock, $\rho_0v_{sh}^{2}$, where $\xi_{CR}$ is the CR acceleration efficiency, $\rho_0$ is the initial density mass (in our model we have assumed a totally ionized medium at the shock), and $v_{sh}$ is the shock velocity. The normalization of electron distribution, $k_{e}$, is then fixed by assuming a standard CR electron/proton ratio in a strong shock, $k_{ep}\approx10^{-2}$. Finally, $p_{inj}$ is the injection momentum. As in the case of SNR shocks, even at the FS of strong stellar wind the injection mechanism is not well known \citep{Casse80}; consequently, we used the same convention used in \cite{Caprioli14_I}, which provides $E_{inj}\sim4.5\frac{1}{2}m_{p}v_{sh}^{2}$.\\
In order to obtain the observed spectrum, we need to consider energy losses affecting the re-accelerated spectrum (see Eq. 15 in Ca16). Protons are affected by ionization losses (at lower energies) and pp-interaction losses (at higher energies) whereas primary and secondary electrons are affected mainly by Bremsstrahlung, Synchrotron, and IC losses (at higher energies) and ionization losses (at lower energies).

To be effective, the energization mechanism at the shock must take place in a time shorter than the minimum value between loss time $t_{loss}$ and interaction time $t_{int}$:
%*******************
\begin{equation}
 t_{acc}\approx D(p)/v_{sh}^{2}< t_{min}=min(t_{int},t_{losses})
 \label{eq:EM condition}
,\end{equation}
%*******************
where $D(E)=\frac{1}{3}r_{L}v(p)\left(\frac{L_c}{r_L}\right)^\delta$ is the diffusion coefficient, strictly related to the particle Larmor radius, $r_{L}=\frac{mv}{eB_{0}}$, to particle velocity $v(p)$, to the perturbation spectrum and, consequently, to the correlation length of magnetic field perturbations, $L_{c}$. From this condition, we obtain the following relation for the maximum momentum reachable by particles :
%************************
%\begingroup\makeatletter\def\f@size{12}\check@mathfonts
\begin{equation}
p_{\rm max}\propto \left(B_{0}\right)\left(v_{sh}\right)^{\frac{2}{1-\delta}}\left(t_{min}\right)^{\frac{1}{1-\delta}}\left(L_{c}\right)^{-\frac{\delta}{1-\delta}}\ ,
\label{Eq:EmaxKo}
\end{equation}
%\endgroup
%*************************
where the magnetic power spectrum index, $\delta$, depends on the turbulence model considered.

%%%%%%%%%%%%%%%%%%%%%%%%%%%%%%%%%%%%%%%
\section{Results}
\label{Sec:Results}
%%%%%%%%%%%%%%%%%%%%%%%%%%%%%%%%%%%%%%%

In this section we illustrate our results in terms of re-acceleration and the possible contribution of acceleration. Our main assumption is that the medium in correspondence with the shock is totally ionized. So far, no radio emission was detected from our ROI, implying that our model has to account for a low synchrotron flux. Also at TeV energies, there is no \gray detection; our computation should therefore return a negligible amount of high energy emission. The number of degrees of freedom of the system can be reduced by considering the known parameters of \kori: its age, $t_{age}\sim7\times10^{6}$ yrs {\citep{Pillitteri16}}; its distance, $d\sim280$ pc {\citep{Pillitteri16}}; and the radius of the surrounding ring of diffuse emission, $R\sim6$ pc. This radius is an intermediate value between the one estimated for the \gray emission in M18 and the one estimated for the CO shell \citep{Pillitteri16}, because we are considering their interaction region as the location of energization. Considering the age of \kori, we can deduce that this OB star is now in its `snowplow' expansion phase and, consequently, the FS velocity has to be of about $10$ km/s \citep{Lamers99, Pillitteri16}.

To estimate the average particle density corresponding to the \gray excess region, we used HI and H$_2$ column densities from, respectively, the Leiden/Argentine/Bonn (LAB) 21-cm survey \citep{Kalberla05,Hartmann97} and the CO survey described in \cite{Dame01}. The latter was then converted into H$_2$ abundance using the X$_\mathrm{CO}$ conversion factor found in M18. These column densities have been translated into volumetric particle densities by assuming an extension of the region parallel to the line of sight similar to the perpendicular one. The value that we obtained is $n_{0}\sim30$ cm$^{-3}$. Since the estimated density value for the ROI is higher than the ISM one, a large range of correlation length and initial magnetic field values can be taken into account. We know that, in the ISM, $L_{c}\sim100$ pc but in high density medium it can be very small, down to $L_{c}=0.1-0.01$ pc \citep{Houde09}. The initial magnetic field, $B_{0}=b\sqrt{n_0}$ $\mu$G, depends on the parameter $b=\frac{V_{A}}{1.84km/s}$, where $V_{A}$ is the Alfve\'n velocity, and it is equal to 1 in the ISM but can vary between 0.3 and 3 in a high density medium. 

From \cite{Pillitteri16}, we know that the dynamical age of the star formation shell is lower than the \kori age. This piece of information is used to constrain the interaction time (which refers to the starting time of \gray emission production and, therefore, of the energization process) to a value of $t_{int}\sim0.1 t_{age}$. Since the emissivity has to be integrated over the emission volume, another basic parameter for our model is the filling factor $f_V$: assuming a uniform emissivity, the volume of the \kori shell covered by \gray emission is $V=\frac{4}{3}\pi f_S R_{sh}^{3}$, where $R_{sh}$ is the shock radius. We fix $f_V$ at 20$\%$ of the \kori wind shell volume, which seems a reasonable assumption given the extension of the detected \gray emission.

Finally, in order to compute the maximum momentum of the system, we need to consider a specific turbulence spectrum; we used the more conservative Kolmogorov spectrum that provides $\delta=\frac{2}{3}$. In this way, we obtain the  explicit equation for the maximum momentum
%************************ Maximum momentum Kolmogorov
\begingroup\makeatletter\def\f@size{8}\check@mathfonts
\begin{equation}
p_{\rm max}\sim41.8\,GeV/c\,\left(\frac{B_{0}}{15\,\mu G}\right)\left(\frac{v_{sh}}{10 \,km/s}\right)^{6}\left(\frac{t_{min}}{700000\,yrs}\right)^{3}\left(\frac{L_{c}}{0.01\,pc}\right)^{-2}\
\label{Eq:EmaxKo}
,\end{equation}
\endgroup
%*************************
where the normalization values are of the order of the ones used (or estimated) in our best model. Looking at the numerical value of this equation, it is clear that we expect a cut-off at low energies, excluding the possibility of TeV emission from this region.

The last but not the least important issue is our assumption of the presence of a thin compressed radiative shell in correspondence with the interaction region. The physics of the FS \citep{Lamers99} and the presence of the CO shell support the existence of this cooling shell that is the fundamental ingredient for the `crushed cloud' model of \cite{Blandford82}, providing a further compression that enhances the final density, and, consequently, the \gray emission (details in Ca16). However, we will see that the high interaction time counteracts the increment in the \gray flux because of the large amount of energy losses.

%-------------------------------
\subsection{Re-acceleration}
\label{Sec:re-acceleration}
%--------------------------------------------

\begin{figure}[!h]
   \centering 
   \includegraphics[scale=0.5]{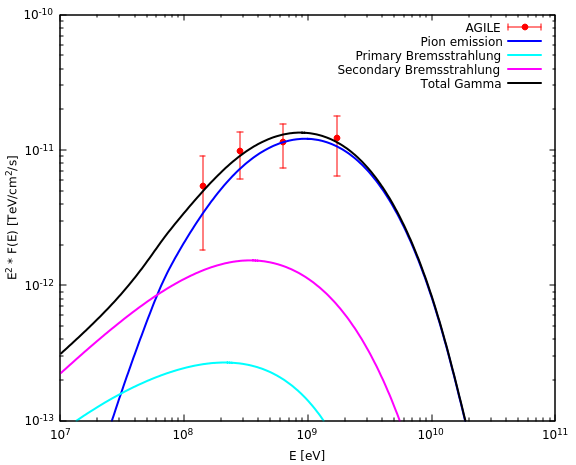}
   \caption{AGILE (red) \gray points plotted with \gray emission from pion decay (blue line), from primary Bremsstrahlung (cyan dashed line), from secondary Bremsstrahlung (magenta dashed line), and total emission (black line).}
   \label{Fig:re-acceleration1}
  \end{figure}

Our best model is based on the assumption that only re-acceleration of pre-existing cosmic rays is present. Diffuse CR protons, He-nuclei, and electrons in the ISM are re-accelerated at the shock between the \kori wind and the near star formation region.\\  
Keeping fixed $n_{0}=30$ cm$^{-3}$ , $v_{sh}=12$ km/s, and $f_V=20\%$, we tuned interaction time, magnetic field parameter $b,$ and correlation length values in order to fit the spectral normalization. Using $t_{int}=0.08t_{age}=5.6\times10^5$ years, $b=2$ (corresponding to $B_{0}=11$ $\mu$G), and $L_{c}=0.01$ pc, we fit the AGILE spectral points remarkably well, obtaining a compressed density $n_{m}= 170$ cm$^{-3}$, a compressed magnetic field $B_{m}=\sqrt{\frac{2}{3}}\frac{n_{m}}{n_{0}}B_0= 50.4$ $\mu$G, and $E_{M}= 22.3$ GeV/n (once again, He nuclei are considered too). In this model, we used the more conservative momentum spectral index used in the linear DSA, $\alpha=4$, which leads to a compression ratio $r_{sh}= 4$. However, a cut-off at such a low energy, together with loss-limited energization, implies a certain degree of covariance in the spectral index values. In fact, models developed with a spectral index in the range $3.5<\alpha<5$ turn out to be equally physically consistent, with only small variations of the other parameters involved. In Fig. \ref{Fig:re-acceleration1} we show the Spectral Energy Distribution (SED) resulting from our best-fit re-acceleration model.

 %^^^^^^^^^^^^^^^^^^^^^^^^^^^^^^^^^^^^^^^^^^^^^^^^^^^^^^^^ Fig. hadronic distributions ^^^^^^^^^^^^^^^^^^^^^^^^^^^^^^^^^^
 %^^^^^^^^^^^^^^^^^^^^^^^^^^^^^^^^^^^^^^^^^^^^^^^^^^^^^^^^ end Fig. hadronic distributions ^^^^^^^^^^^^^^^^^^^^^^^^^^^^^^^^^^

Most of the contribution to \gray emission comes from pion decay; the flux decrement at a few GeV is directly correlated with the particle energy cut-off at about $E_M\sim23$ GeV/n. A particle cut-off at these energies is in agreement with no detection of the Orion region at the TeV energy band. Secondary electrons, produced in inelastic pp-collisions from the decay of charged pions, dominate over primary electrons because of high density and energy losses influence, giving a contribution at $E<200$ MeV through Bremsstrahlung emission. Energy losses are also the reason for very low synchrotron emission ($E^2f(E) < 10^{-14}$ TeV/cm$^{-2}$/s, not shown in Fig. \ref{Fig:re-acceleration1}). Our prediction concerning radio emission is in agreement with the lack of radio detection in the ROI.

We want to stress that the contribution of ``crushed cloud'' adiabatic compression is not really necessary in order to explain the \gray emission detected by AGILE. Using the same parameters, we also have a good fit  with only re-accelerated particles with different final values because of the lack of adiabatic compression: $n_{m}= 120$ cm$^{-3}$, a compressed magnetic field $B_{m}= 35.8$ $\mu$G, and $E_{M}= 51.4$ GeV/n (we also consider He nuclei). This result is important because the diffusion length inside the thin shell in a time $t_{int}$, is $L_{diff}=\sqrt{D(E)t_{int}}\sim4.2$ pc, which is a value lower than the shock radius $R_{sh}$, but likely greater than the interaction region scale. This means that it could be very difficult to confine re-accelerated particles inside the radiative shell for the time $t_{int}$ and compress them adiabatically. In such a scenario, we verified that the downstream re-accelerated particles alone could make a sufficient contribution to account for the Orion region \gray excess.
 
%------------------------------
\subsection{Acceleration contribution}
\label{Sec:acceleration}
%------------------------------
We showed that the contribution of re-accelerated particles (with or without the adiabatic compression) is sufficient to fully explain the \gray excess seen by AGILE with reasonable physical parameters. However, we tested the hypothesis of a possible further contribution from freshly accelerated particles.

\begin{figure}[!h]
   \centering 
   \includegraphics[scale=0.5]{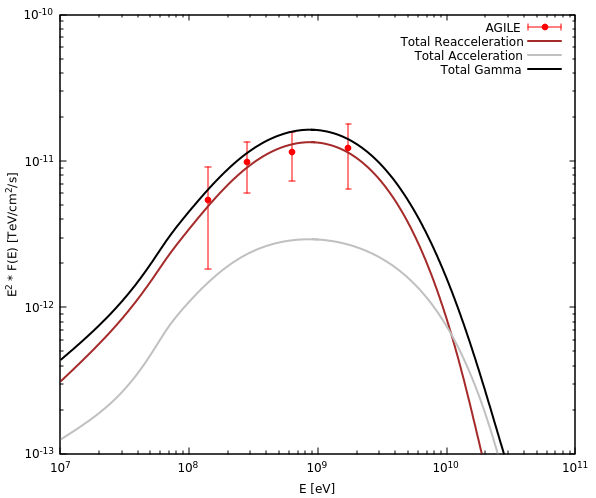}
   \caption{AGILE (red) \gray points plotted with the total \gray emission produced by re-acceleration (brown line) and acceleration (grey line, $\xi_{CR}=20\%$) described with parameters of our best model. The black line is the sum of the two contributions.}
   \label{Fig:acc+Reacc}
  \end{figure}

Keeping the physical parameters of our best re-acceleration model, we calculated an upper limit to the acceleration efficiency $\xi_{CR}$ such that the total \gray emission, due to both re-acceleration and acceleration, is compatible with the AGILE spectrum. The value we found, $\xi_{CR}\sim 6 \%$, shows that a contribution from freshly accelerated CRs, although plausible, would anyway be marginal compared to the one from GCR re-acceleration, as shown in Fig. \ref{Fig:acc+Reacc}.

%^^^^^^^^^^^^^^^^^^^^^^^^^^^^^^^^^^^^^^^^^^^^^^^^^^^^^^^^ Fig. acc+reacc distributions ^^^^^^^^^^^^^^^^^^^^^^^^^^^^^^^^^^
 %^^^^^^^^^^^^^^^^^^^^^^^^^^^^^^^^^^^^^^^^^^^^^^^^^^^^^^^^ end Fig. acc+Reacc distributions ^^^^^^^^^^^^^^^^^^^^^^^^^^^^^^^^^^

The grey line in Fig. \ref{Fig:acc+Reacc} represents the total contribution from acceleration emission process, primary and secondary Bremsstrahlung, and pion emission. Given the low velocity of the \kori FS and the long interaction time, the low contribution from accelerated particles is not surprising. Galactic CRs are present everywhere in the Galaxy and, if there are energization conditions, for instance, at the SNR shocks or at the stellar wind shocks, they are always re-accelerated. The present case resembles that of the middle-aged SNR W44 (see Ca16), for which the \gray flux is explained in terms of a dominant re-acceleration process. A dominant contribution from freshly accelerated particles, instead, can be found in young objects, such as young SNRs or stellar wind TS; these are characterized by shocks with very high velocities (order of $10^{3}$ km/s), for which strong \gray fluxes can be achieved with reasonable values of CR efficiency.

%------------------------------
\subsection{Compression only}
\label{Sec:only_Comp}
%------------------------------

\begin{figure}[!h]
   \centering 
   \includegraphics[scale=0.5]{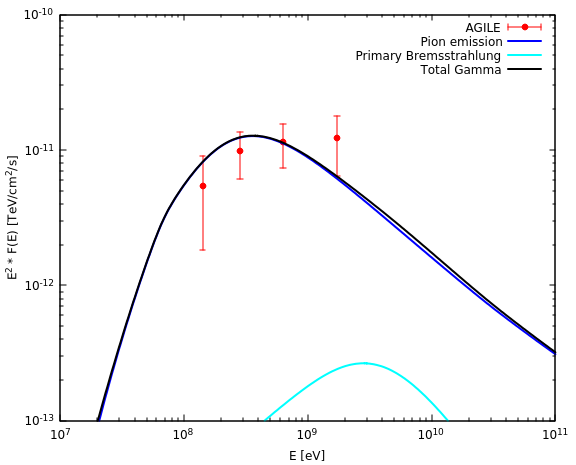}
   \caption{Same data and model curves as Fig. \ref{Fig:re-acceleration1}, assuming that only adiabatic compression is present.}
   \label{Fig:only_comp}
  \end{figure}

One of the main assumptions of our model is that the radiative shell is totally ionized. This condition allows the development of a strong turbulence that energizes CR particles. However, since the presence of neutrals could efficiently dampen the turbulence \citep{Chevalier99,Bykov00}, we computed a lower limit for their density found by the condition that the ion-neutral damping time is longer than the interaction time. Following \cite{Ptuskin03}, the ion-neutral damping is
\begin{equation}
\Gamma_{\rm IN}=
\left\{
\begin{array}{cc}
\frac{k^2V_{A}^{2}}{\nu_{IN}\left(1+\frac{n_i}{n_H}\right)^2} & k<k_c\\
\frac{\nu_{\rm IN}}{2} & k>k_c
\end{array}
\right.
,\end{equation}
where $\nu_{\rm IN}=8.4\times10^{-9}{\rm s}^{-1}\ \left(T/10^{4}K\right)^{0.4}n_{H}$ is the ion-neutral collision frequency, with $T$ the temperature, and $n_H$ the density of neutrals in units of ${\rm cm}^{-3}$. The critical wave number separating the two regimes is $k_c=(\nu_{\rm IN}/V_A)(n_i/n_H)$, which corresponds to perturbations of wavelength resonant with particles of energy $E_c\approx 1.7\times 10^5$ GeV in the regime of density and magnetic field strength we are considering. Consequently, we consider the ion-neutral damping rate for $E<E_c$ as corresponding to $k<k_c$, and by requiring that $\Gamma_{IN}<1/t_{int}$ we obtain a lower limit for the neutral fraction, $n_{H}<1.3\times10^{-5}$ cm$^{-3}$ , which is a very stringent condition. Even considering the self-generated turbulence \citep{Cardillo16}, we cannot enhance this lower limit; consequently, we try to explain the AGILE-detected \gray emission using only adiabatic compression, without any energization. This model is shown in Fig. \ref{Fig:only_comp}.

%^^^^^^^^^^^^^^^^^^^^^^^^^^^^^^^^^^^^^^^^^^^^^^^^^^^^^^^^ Fig. compression only distributions ^^^^^^^^^^^^^^^^^^^^^^^^^^^^^^^^^^
 %^^^^^^^^^^^^^^^^^^^^^^^^^^^^^^^^^^^^^^^^^^^^^^^^^^^^^^^^ end Fig. acc+Reacc distributions ^^^^^^^^^^^^^^^^^^^^^^^^^^^^^^^^^
The parameters of this model are the same as our re-acceleration one but here the interaction time is set to be of the order of $t_{int}=10^{-4}t_{age}=700$ yrs. This lower value can be understood because a lower interaction time leads to lower energy losses and a higher \gray emission. However, the available time is too short for intense secondary electron production and secondary Bremsstrahlung contribution is totally negligible. We can say that, even in case of turbulence damping, an adiabatic compression could explain the AGILE \gray emission with the condition that the \kori shock and the CO shell have only been interacting for $10^{3}$ yrs.
A rough estimation of the shell cooling time from the cooling time curve in \cite{Draine11}, using the density value of our model, gives $\tau_{cool}\sim10^{2}$ yrs $< t_{int}$, confirming the plausibility of the only-compression scenario. Moreover, in this model, the diffusion length issue inside the radiative shell discussed in Sect. \ref{Sec:re-acceleration} is automatically solved. In the re-acceleration model, the presence of a compressed shell implies a diffusion length that is too high; assuming instead a purely adiabatic compression, the lower value of the interaction time provides a value $L_{diff}\sim 0.01$ pc, implying a higher chance of confining and compressing particles adiabatically.

%%%%%%%%%%%%%%%%%%%%%%%%%%%%%%%%%%%%%%%
\section{Discussion}
\label{Sec:Discussions}
%%%%%%%%%%%%%%%%%%%%%%%%%%%%%%%%%%%%%%%

It is generally assumed \citep{Casse80, Voelk82, Cesarsky83} that CR acceleration due to strong stellar winds takes place in the TS. However, the asymmetry in the AGILE- and Fermi-LAT-detected \gray emission from the Orion region is not easily understandable in this scenario; a wind TS expands symmetrically into the stellar wind, having the same upstream everywhere in the wind bubble. Therefore, we associated the high energy emission to the interaction of the slower FS with the CO shell described in \cite{Pillitteri16}, which approximately coincides with the AGILE emission region. A more precise measurement of the \kori FS velocity is very important. Small variations of its value have a strong impact on the model: velocities of about 5 km/s or lower would make the proposed mechanism of CR energization virtually impossible, because both the maximum momentum and normalization become too low. Given the sensitivity of the model to the shock velocity, we decided to fix this parameter to a value consistent with the FS physics and that provides a validity range physically consistent for all the other parameters.

From \cite{Pillitteri16} we know that \kori started to sweep up the shell about 1.5 Myr ago; this implies that the interaction time cannot be much lower than the age of \kori ($7$ Myrs). Assuming an interaction time lower than $t_{int}=0.08t_{age}$ by an order of magnitude, particles do not have the necessary time to be re-accelerated and they can only be compressed, providing a \gray emission that cannot fit the AGILE data. If we hypothesized, instead, that the interaction time is of the same order of the age of \kori , energy loss processes would dissipate very large amounts of energy, leading to a \gray flux much lower than the one observed.

Analysing the behaviour of our model as the density changes, we found that the presence of a thin radiative shell is not a fundamental condition in order to have the correct amount of \gray flux. Indeed, excluding the only adiabatic compression, we obtain an equally consistent model. This can be explained by the fact that an emission due to a compressed density that is too high, when associated to a long interaction time, leads to large energy losses with a consequent flux suppression. On the other hand, if the initial density is too low, there are not enough target particles to justify the \gray emission observed by AGILE.% Assuming instead an initial density value of about $100$ cm$^{-3}$, we could obtain a good fit by enhancing the interaction time, the initial magnetic field and the shock velocity (about $18$ $km/s$), thanks to their mutual relations and their affection on losses and maximum momentum. 

On the other hand, we also analysed a scenario in which our assumption of totally ionized shocked material is not real, which implies perturbation damping due to ion-neutral collisions and, consequently, the lack of an effective re-acceleration. We showed that, even with only adiabatic compression within the thin shell, Galactic CRs are energized in a sufficient way to explain the \gray detected emission.

Another important parameter, very difficult to estimate, is the coherence length, which we assumed to be of the order of its estimated minimum value in a high density medium \citep{Houde09}. If we hypothesize that its value is higher by an order of magnitude, that is, $L_C=0.1 pc$, AGILE data can be fitted only by significant modifications of the other model free parameters. In particular, an increase of both shock velocity (about $18$ $km/s$) and filling factor ($f_{V}\sim25\%$) would be needed to compensate for a decrement in the maximum momentum and an increase in the acceleration time (Ca16). Such values, however, seem to be unrealistic in the context described here.

We already mentioned in our previous work  that the extended source data analysis leads to a hard photon spectral index, $1.7 \pm 0.2$. However, in a DSA context the final photon spectral index is the result of the parent proton population distribution spectrum being affected by different processes, such as re-acceleration, energy losses, and diffusion. Consequently, the photon spectral index obtained by a best-fit analysis cannot give stringent constraints on the original proton distribution. In our model, the injection proton index is strictly correlated with the compression ratio of the shock; we found equally consistent models for a large range of its values, [2.5-7], corresponding (in a strong shock) to a range for the momentum spectral index between [5-3.5]. This large range is due to the very low high energy cut-of,f which provides a suppression of the photon flux before we can observe different spectral behaviour due to different proton spectral indices.

Regardless of the combination of parameters used, in our work we can draw two important conclusions. Firstly, even using different realistic values for all relevant parameters, we cannot obtain a maximum momentum higher than about $50$ GeV/n, definitely ruling out high energy emission, in agreement with the absence of detections at TeV energies. Moreover, any model seems to provide a radio emission above $10^{-14}$ TeV/cm2/s, in agreement with the lack of radio detection in our ROI. Another important conclusion we reached is that Galactic CR re-acceleration can fully explain the AGILE and Fermi-LAT \gray excess detected in the Orion region. Freshly accelerated particles only provide a marginal contribution, if any.

%%%%%%%%%%%%%%%%%%%%%%%%%%%%%%%%%%%%%%%
\section{Summary and conclusions}
\label{Sec:Conclusions}
%%%%%%%%%%%%%%%%%%%%%%%%%%%%%%%%%%%%%%%
Because of the large amount of ISM it hosts, the Orion region is a very important target for the study of diffuse \gray emission in our Galaxy. In recent years, the Fermi-LAT \citep{Ackermann12} and AGILE  \citep{Marchili18} satellites allowed us to map this emission in a more precise way and to model it according to a `standard model' that includes pp and Bremsstrahlung emission from atomic and molecular hydrogen ($HI$ and H$_{2}$), isotropic emission from inverse Compton scattering in the ISRF, point-source emission, and extragalactic isotropic emission. An in-depth analysis of both the Fermi-LAT and the AGILE data reveals a significant \gray excess, which cannot be explained by \dg contributions \citep{Grenier05}.

In \cite{Ackermann12} this \gray excess is explained with a non-linearity in the CO-to-H$_{2}$ relation. Here we try to model it in the context of CR energization in the strong stellar wind from the OB star \kori. CR acceleration in the TS of an OB star is a mechanism that is considered a reasonable alternative to acceleration in SNR shocks \citep{Casse80, Voelk82, Cesarsky83}. We developed here an alternative scenario in which particle energization takes place in correspondence with the interaction region between the slow FS of \kori and a dense star formation region, shaped in a dense shell detected in IR and CO maps. It surrounds the detected \gray emission and includes many OB stars, analysed in the X-ray band by \cite{Pillitteri16}. It is likely that this dense shell was swept-up during the \kori stellar wind expansion.

This scenario is very similar to the standard one found in the SNR/MC interaction context. Consequently, in order to model the AGILE \gray excess, we followed the re-acceleration and acceleration model described in \cite{Cardillo16}, considering Galactic CR spectra (for protons, Helium, and electrons) measured from Voyager and PAMELA, primary and secondary electron contributions, and energy losses. The main constraint in our model is the shock velocity. The OB star \kori is now in its `snowplow' expansion phase \citep{Lamers99} and the FS is very slow, with a velocity of about $10$ km/s. Only the combination of a high density target with a quite long interaction time can give the necessary conditions to allow CR energization in such a slow shock.

We estimate the average density in correspondence with the emission region to be of the order of $n_{0}=30$ cm$^{-3}$. The physics of the FS \citep{Lamers99}, together with the interaction with a MC, leads to the formation of a thin compressed radiative shell according to \cite{Blandford82}, implying a further compression of the medium. The interaction time has to be very high, because we know that the star formation shell has been swept-up by the \kori shell for about $10^{6}$ yrs. For this reason, energy losses give the highest contribution in the estimation of the maximum momentum. These considerations allowed us to constrain magnetic field and correlation length values, fixing an emitting volume filling factor of about $20 \%$.

Taking into account the constraints on the FS velocity and on the interaction time, we obtained a solid re-acceleration model, showing how the AGILE \gray excess can be explained by first order Fermi re-acceleration of particles at the shock between \kori wind and the star formation shell. Radio emission provided by this model is very low, in agreement with observations, and secondary Bremsstrahlung emission dominates the primary one, mainly because of energy losses. IC contribution is totally negligible.

Unfortunately, our analysis can give very little information about the parent proton spectral distribution. The best model that we found provides a very low cut-off energy, $E_{M}\sim 23$ GeV/n, implying that we cannot distinguish between different spectral behaviours due to a proton spectral index that is high or steep. We obtain a good fit of the AGILE data with a spectral index in a range between [3.5-5].

In spite of possible variations in some of the assumed parameters and the ambiguity of others, we can fix some important constraints. Even if a contribution from freshly accelerated particles is present, it is always marginal with respect to the re-acceleration one and this is mainly due to the very low FS velocity. Moreover, our re-acceleration model provides negligible \gray emission beyond energies of about $E_{M}\sim50$ GeV/n, and a radio flux lower than about $10^{-14}$ TeV/cm$^{2}$/s, in agreement with the lack of TeV and radio detection in the ROI. The AGILE \gray excess from the Orion region could be the first detection of a FS stellar wind.

Moreover, we showed that, even relaxing our hypothesis of a completely ionized medium and, consequently, considering a possible inhibition of re-acceleration due to ion-neutral collisions, we can explain the AGILE \gray detection from the Orion region invoking only adiabatic compression, without re-acceleration. In this case, a lower interaction time, $t_{int}=700$ yrs, is required.
For a more precise characterization of the emitting region, multi-wavelength analysis of the Orion region, focused on the \gray excess detected by AGILE and Fermi-LAT, is required.

\begin{acknowledgements}
We wish to thank the anonymous referee for the useful comments which significantly improved the paper.
\end{acknowledgements}

\end{document}